\documentclass[12pt,aps,prd,showpacs,amsmath,amssymb]{revtex4}
\input epsf
\begin{document}
\title{Semileptonic Decays of $B$ Meson Transition Into $D$-wave Charmed Meson Doublets}
\author{Long-Fei Gan}
 \email{lfgan@nudt.edu.cn}
\author{Ming-Qiu Huang}
\affiliation{Department of Physics, National University of Defense
Technology, Hunan 410073, China}
\date{\today}
\begin{abstract}
We use QCD sum rules to estimate the leading-order universal form
factors describing the semileptonic $B$ decay into orbital excited
$D$-wave charmed doublets, including the ($1^{-}$, $2^{-}$) states
($D_{1}^{*}$, $D'_{2}$) and the ($2^{-}$, $3^{-}$) states ($D_{2}$,
$D_{3}^{*}$). The decay rates we predict are $\Gamma_{B\rightarrow
D^{*}_{1}\ell\overline{\nu}}=\Gamma_{B\rightarrow
D'_{2}\ell\overline{\nu}}=2.4\times10^{-18} \mbox{GeV}$,
$\Gamma_{B\rightarrow D_{2}\ell\overline{\nu}}=6.2\times10^{-17}
\mbox{GeV}$, and $\Gamma_{B\rightarrow
D_{3}^{*}\ell\overline{\nu}}=8.6\times10^{-17} \mbox{GeV}$. The
branching ratios are $\mathcal {B}(B\rightarrow
D^{*}_{1}\ell\overline{\nu})=\mathcal {B}(B\rightarrow
D'_{2}\ell\overline{\nu})=6.0\times10^{-6}$, $\mathcal
{B}(B\rightarrow D_{2}\ell\overline{\nu})=1.5\times10^{-4}$, and
$\mathcal {B}(B\rightarrow
D_{3}^{*}\ell\overline{\nu})=2.1\times10^{-4}$, respectively.
\end{abstract}
 \maketitle

\section{Introduction}\label{sec1}
Higher excitations than $D^{(*)}$ play an important role in the
understanding of semileptonic $B$ decays. Knowledge of these
processes is important to reduce the uncertainties of the
measurements on other semileptonic $B$ decays, and thus the
determination of the Cabibbo-Kobayashi-Maskawa matrix elements, such
as $|V_{cb}|$. Theoretically, the semileptonic decay processes are
described by some form factors. The challenge for theory is the
calculation of these decay form factors. Fortunately, the heavy
quark effective theory (HQET) \cite{Wis}, with an expansion in terms
of $1/m_{Q}$ for hadrons containing a single heavy quark, provides a
systematic method for investigating such processes. In HQET the
approximate symmetries allow one to organize the spectrum of heavy
mesons according to parity $P$ and total angular momentum $s_{l}$ of
the light degree of freedom. Coupling the spin of the light degrees
of freedom $s_{l}$ with the spin of a heavy quark $s_{Q}=1/2$ yields
a doublet of meson states with a total spin $s=s_{l}\pm1/2$. For
charmed mesons, the lowest lying states $(0^{-}, 1^{-})$ doublet
($D$, $D^{*}$) are $S$-wave states with the spin of light degrees
$s_{l}=1/2$. The $P$-wave excitation corresponds to two series of
states, one is the $s_{l}=1/2$ series, the $(0^{+}, 1^{+})$ doublet
($D_{0}^{*}$, $D'_{1}$); the other is the $s_{l}=3/2$ series, the
$(1^{+}, 2^{+})$ doublet ($D_{1}$, $D^{*}_{2}$). For $D$-wave
states, those are $(1^{-}, 2^{-})$ and $(2^{-}, 3^{-})$ doublets
(($D_{1}^{*}$, $D'_{2}$) and ($D_{2}$, $D^{*}_{3}$)), corresponding
to the spin of light degrees of freedom $s_{l}=3/2$ and $s_{l}=5/2$.
The early study of the heavy-light mesons can be found in Ref.
\cite{God}. The $S$-wave and $P$-wave charmed states have been
observed so far. The properties of these states have been
extensively studied using different approaches during the past few
years, including masses \cite{Ebe,Dai}, decay constants
\cite{Neu,Cap,Cve}, and decay widths \cite{Dai3,Dai2,Eic,Xia}. For
the $D$-wave charmed mesons, their properties were investigated with
the potential model \cite{Eic} and QCD sum rules \cite{Wei}.

Semileptonic $B$ decay into an excited heavy meson has been observed
in experiments \cite{Cle,Ale}. Recently, BABAR has measured
semileptonic $B$ decays into orbitally excited charmed mesons
$D_{1}(2420)$ and $D^{*}_{2}(2460)$ \cite{Aub1}. They also reported
two new $D_{s}$ states $D_{sJ}(2860)$ and $D_{sJ}(2690)$ in the $DK$
channel, which may fit in the $D$-wave charm-strange doublets
\cite{Aub}. A similar state $D_{sJ}(2715)$ has also been observed by
Belle \cite{Bel}. It is expected that the nonstrange $D$-wave
charmed mesons will be found, and the measurements of the
semileptonic $B$ decays into these states become available in the
near future. To this end we study the predictions of HQET for
semileptonic $B$ decays to $D$-wave charmed mesons.

The semileptonic decay rate of a B meson transition into an charmed
meson is determined by the corresponding matrix elements of the weak
axial-vector and vector currents. In the heavy quark limit these
elements are described, respectively, by one universal Isgur-Wise
function at the leading order of heavy quark expansion \cite{Wis1}.
The universal Isgur-Wise function is a nonperturbtive parameter. It
must be calculated in some nonperturbative approaches. The main
theoretical approaches are QCD sum rules \cite{Shi}, constituent
quark models, and lattice QCD. The investigations of semileptonic
$B$ decays into charmed mesons can be found in Refs.
\cite{Neu,Hua1,Ebe1,Dea,Wis1} with different methods. In this work,
we estimate the leading-order Isgur-Wise functions describing the
decays $B\rightarrow (D_{1}^{*}, D'_{2})\ell\overline{\nu}$ and
$B\rightarrow (D_{2}, D_{3}^{*})\ell\overline{\nu}$ and give a
prediction for the widths of the decays.

The remainder of this paper is organized as follows. In Sec.
\ref{sec2} we present the formulas of weak current matrix elements
and decay rates. In Sec. \ref{sec3} we give the relevant sum rules
for two-point correlators, and then deduce the three-point sum rules
for the Isgur-Wise functions. Section \ref{sec4} is devoted to
numerical results and discussions.

\section{Analytic formulations for semileptonic decay
amplitudes $B\rightarrow (D_{1}^{*}, D'_{2})\ell\overline{\nu}$ and
$B\rightarrow (D_{2}, D_{3}^{*})\ell\overline{\nu}$} \label{sec2}
The heavy-light meson doublets can be expressed conveniently by
effective operators \cite{Fal}. For the ground doublet, the operator
is
\begin{equation}\label{operator1}
 H_{a}=\frac{1+\rlap/v}{2}[D^{*}_{\mu}\gamma^{\mu}-D\gamma_{5}].
 \end{equation}
The effective operators describing the meson doublets
$D(1^{-},2^{-})$ and $D(2^{-},3^{-})$ are given by
\begin{equation}\label{operator2}
 X^{\mu}=\frac{1+\rlap/v}{2}[D'^{\mu\nu}_{2}\gamma_{5}\gamma_{\nu}-D^{*}_{1\nu}\sqrt{\frac{3}{2}}(g^{\mu\nu}-\frac{1}{3}\gamma^{\nu}(\gamma^{\mu}+v^{\mu}))],
 \end{equation}
and
\begin{equation}\label{operator3}
 H^{\mu\nu}=\frac{1+\rlap/v}{2}[D^{*\mu\nu\sigma}_{3}\gamma_{\sigma}-\sqrt{\frac{3}{5}}\gamma_{5}
 D^{\alpha\beta}_{2}(g^{\mu}_{\alpha}g^{\nu}_{\beta}-\frac{\gamma_{\alpha}}{5}g^{\nu}_{\beta}
 (\gamma^{\mu}-v^{\mu})-\frac{\gamma_{\beta}}{5}g^{\mu}_{\alpha}
 (\gamma^{\nu}-v^{\nu}))].
 \end{equation}
In these operators, $D^{*}_{\mu}$, $D$, $D'^{\mu\nu}_{2}$,
$D^{*}_{1\nu}$, $D^{*\mu\nu\sigma}_{3}$, and $D^{\alpha\beta}_{2}$
separately represent annihilation operators of the $Q\overline{q}$
mesons with appropriate quantum numbers and $\rlap/v =
v\cdot\gamma$, $v$ is the heavy meson velocity. The theoretical
description of semileptonic decays involves the matrix elements of
vector and axial-vector currents
($V^{\mu}=\overline{c}\gamma^{\mu}b$ and
$A^{\mu}=\overline{c}\gamma^{\mu}\gamma_{5}b$) between $B$ mesons
and excited $D$ mesons. For the processes $B\rightarrow (D^{*}_{1},
D'_{2})\ell\overline{\nu}$ and $B\rightarrow (D_{2},
D_{3}^{*})\ell\overline{\nu}$, these matrix elements can be
parametrized through applying the trace formalism as follows
\cite{Fal}:
\begin{eqnarray}\label{matrix1}
\langle D^{*}_{1}(v^{'},\varepsilon)|(V-A)^{\mu}|B(v)\rangle&=&
\sqrt{\frac{3}{2}}\sqrt{m_{B}m_{D^{*}_{1}}}\tau_{1}(y)[\varepsilon^{*}\cdot
v(v^{\mu}-\frac{y+2}{3}v'^{\mu})\nonumber\\&-&i\frac{1-y}{3}\epsilon^{\mu\alpha\beta\sigma}\varepsilon^{*}_{\alpha}v'_{\beta}v_{\sigma}],
\end{eqnarray}
\begin{equation}\label{matrix2}
\langle D'_{2}(v^{'},\varepsilon)|(V-A)^{\mu}|B(v)\rangle =
\sqrt{m_{B}m_{D'_{2}}} \tau_{1}(y)
\varepsilon^{*}_{\rho\nu}v^{\rho}[g^{\mu\nu}(y-1)-v^{\nu}v^{'\mu}
+i\epsilon^{\alpha\beta\nu\mu}v^{'}_{\alpha}v_{\beta}],
\end{equation}
and
\begin{eqnarray}\label{matrix3}
\langle D_{2}(v^{'},\varepsilon)|(V-A)^{\mu}|B(v)\rangle&=&
\sqrt{\frac{5}{3}}\sqrt{m_{B}m_{D_{2}}}\tau_{2}(y)
\varepsilon^{*}_{\alpha\beta}v^{\alpha}[\frac{2(1-y^{2})}{5}g^{\mu\beta}-v^{\beta}v^{\mu}
+\frac{2y-3}{5}v^{\beta}v^{'\mu}\nonumber\\&+&i\frac{2(1+y)}{5}\epsilon^{\mu\lambda\beta\rho}v_{\lambda}v^{'}_{\rho}],
\end{eqnarray}
\begin{equation}\label{matrix4}
\langle D_{3}^{*}(v^{'},\varepsilon)|(V-A)^{\mu}|B(v)\rangle =
\sqrt{m_{B}m_{D_{3}^{*}}} \tau_{2}(y)
\varepsilon^{*}_{\alpha\beta\lambda}v^{\alpha}v^{\beta}[g^{\mu\lambda}(1+y)-v^{\lambda}v^{'\mu}
+i\epsilon^{\mu\lambda\rho\tau}v_{\rho}v^{'}_{\tau}],
\end{equation}
where $(V-A)^{\mu}=\overline{c}\gamma^{\mu}(1-\gamma_{5})b$ is the
weak current, $y=v\cdot v^{'}$ and $\tau_{1}(y)$, $\tau_{2}(y)$ are
the universal form factors, and $\varepsilon^{*}_{\alpha}$,
$\varepsilon^{*}_{\alpha\beta}$,
$\varepsilon^{*}_{\alpha\beta\lambda}$ are the polarization tensors
of these mesons. The differential decay rates are calculated by
making use of the formulas (\ref{matrix1}) to (\ref{matrix4}) given
above:
\begin{equation}\label{rate1}
\frac{d\Gamma}{dy}(B\rightarrow D^{*}_{1}\ell\overline{\nu})=
\frac{G^{2}_{F}V^{2}_{cb}m^{2}_{B}m^{3}_{D^{*}_{1}}}{72\pi^{3}}(\tau_{1}(y))^{2}(y-1)^{\frac{5}{2}}
(y+1)^{\frac{3}{2}}[(1+r_{1}^{2})(2y+1)-2r_{1}(y^{2}+y+1)],
\end{equation}
\begin{equation}\label{rate2}
\frac{d\Gamma}{dy}(B\rightarrow D'_{2}\ell\overline{\nu})=
\frac{G^{2}_{F}V^{2}_{cb}m^{2}_{B}m^{3}_{D'_{2}}}{72\pi^{3}}(\tau_{1}(y))^{2}(y-1)^{\frac{5}{2}}
(y+1)^{\frac{3}{2}}[(1+r_{2}^{2})(4y-1)-2r_{2}(3y^{2}-y+1)],
\end{equation}
\begin{equation}\label{rate3}
\frac{d\Gamma}{dy}(B\rightarrow D_{2}\ell\overline{\nu})=
\frac{G^{2}_{F}V^{2}_{cb}m^{2}_{B}m^{3}_{D_{2}}}{360\pi^{3}}(\tau_{2}(y))^{2}(y-1)^{\frac{5}{2}}
(y+1)^{\frac{7}{2}}[(1+r_{3}^{2})(7y-3)-2r_{3}(4y^{2}-3y+3)],
\end{equation}
\begin{equation}\label{rate4}
\frac{d\Gamma}{dy}(B\rightarrow D_{3}^{*}\ell\overline{\nu})=
\frac{G^{2}_{F}V^{2}_{cb}m^{2}_{B}m^{3}_{D_{3}^{*}}}{360\pi^{3}}(\tau_{2}(y))^{2}(y-1)^{\frac{5}{2}}
(y+1)^{\frac{7}{2}}[(1+r_{4}^{2})(11y+3)-2r_{4}(8y^{2}+3y+3)],
\end{equation}
with $r_{i}=\frac{m_{D_{i}}}{m_{B}}$ ($D_{i}=D^{*}_{1}, D'_{2},
D_{2}, D^{*}_{3}$ for $i=1, 2, 3, 4$ ). In the equations above, we
have presented the decay rates of B semileptonic decay processes
$B\rightarrow (D^{*}_{1}, D'_{2})\ell\overline{\nu}$ and
$B\rightarrow (D_{2}, D^{*}_{3})\ell\overline{\nu}$ in terms of the
universal form factors $\tau_{1}(y)$ and $\tau_{2}(y)$,
respectively. The only unknown factors in these equations are
$\tau_{1}(y)$ and $\tau_{2}(y)$, which need to be determined by
nonperturbative methods.

\section{Sum rules for Isgur-Wise functions}
\label{sec3}In the calculation of Isgur-Wise functions in HQET by
means of QCD sum rule, the interpolating currents are potentially
important. In Ref. \cite{Dai}, two series of interpolating currents
with nice propertties were proposed:
\begin {equation}\label{current1 }
J^{\dag\alpha_{1}\ldots\alpha_{j}}_{j,P,i}=\overline{h}_{v}(x)
\Gamma^{\{\alpha_{1}\ldots\alpha_{j}\}}_{j,P,i}(D_{x_{t}})q(x)
\end{equation}
 or
\begin {equation}\label{current2 }
J^{'\dag\alpha_{1}\ldots\alpha_{j}}_{j,P,i}=\overline{h}_{v}(x)
\Gamma^{\{\alpha_{1}\ldots\alpha_{j}\}}_{j,P,i}(D_{x_{t}})(-i)\rlap/
D_{x_{t}}q(x)
\end{equation}
where $i=1,2$ corresponding to two series of doublets of the
spin-parity $[j^{(-1)^{j+1}},(j+1)^{(-1)^{j+1}}]$ and
$[j^{(-1)^{j}},(j+1)^{(-1)^{j}}]$, respectively.
$D_{t\mu}=D_{\mu}-v_{\mu}(v\cdot D)$ is the transverse component of
the covariant derivative with respect to the velocity of the meson
and
\begin{equation}\label{semmtrize}
\Gamma^{\{\alpha_{1}\ldots\alpha_{j}\}}(D_{x_{t}})=\text{symmetrize}
\{\Gamma^{\alpha_{1}\ldots\alpha_{j}}(D_{x_{t}})-\frac{1}{3}g^{\alpha_{1}\alpha_{2}}_{t}
g^{t}_{\alpha^{'}_{1}\alpha^{'}_{2}}\Gamma^{\alpha^{'}_{1}\alpha^{'}_{2}\alpha_{3}\cdots\alpha_{j}}\}
\end{equation}
with the transverse metric $g^{\alpha\beta}_{t}=g^{\alpha\beta}-
v^{\alpha}v^{\beta}$. For the doublets of spin-parity
$[j^{(-1)^{j+1}},(j+1)^{(-1)^{j+1}}]$ and
$[j^{(-1)^{j}},(j+1)^{(-1)^{j}}]$, the expressions for
$\Gamma^{\alpha_{1}\ldots\alpha_{j}}(D_{x_{t}})$ have been
explicitly given in \cite{Dai} as
\[
\Gamma(D_{x_{t}})=\left\{\begin{tabular}{ll}
$\sqrt{\frac{2j+1}{2j+2}}\gamma^{5}(-i)^{j}D^{\alpha_{2}}_{x_{t}}\cdots
D^{\alpha_{j}}_{x_{t}}
(D^{\alpha_{1}}_{x_{t}}-\frac{j}{2j+1}\gamma^{\alpha_{1}}_{t}\rlap/D_{x_{t}}) $, & for $j^{(-1)^{j+1}}$\\
$\frac{1}{\sqrt{2}}\gamma^{\alpha_{1}}_{t}(-i)^{j}D^{\alpha_{2}}_{x_{t}}\cdots
D^{\alpha_{j}}_{x_{t}}$, & for $(j+1)^{(-1)^{j+1}}$
\end{tabular}
\right.
\]
\[
\Gamma(D_{x_{t}})=\left\{\begin{tabular}{ll}\
$\frac{1}{\sqrt{2}}\gamma^{5}(-i)^{j}\gamma^{\alpha_{1}}_{t}D^{\alpha_{2}}_{x_{t}}\cdots
D^{\alpha_{j+1}}_{x_{t}} $, & for $(j+1)^{(-1)^{j}}$\\
$\sqrt{\frac{2j+1}{2j+2}}(-i)^{j}D^{\alpha_{2}}_{x_{t}}\cdots
D^{\alpha_{j}}_{x_{t}}(D^{\alpha_{1}}_{x_{t}}-\frac{j}{2j+1}\gamma^{\alpha_{1}}_{t}\rlap/
D_{x_{t}}) $, & for $j^{(-1)^{j}}$
\end{tabular}
\right.
\]
where $\gamma_{t\mu}=\gamma_{\mu}-\rlap/vv_{\mu}$ is the transverse
component of $\gamma_{\mu}$ with respect to the heavy quark
velocity.

For the $D$-wave meson doublets with $s_{l}=\frac{3}{2}^{-}$ and
$s_{l}=\frac{5}{2}^{-}$, where $j=1$ and $j=2$, the currents are
given by the following expressions:
\begin{equation}\label{current3}
J^{\dag\alpha}_{1,-,3/2}=-i\sqrt{\frac{3}{4}}\overline{h}_{v}(D^{\alpha}_{t}-\frac{1}{3}\gamma^{\alpha}_{t}\rlap/D_{t})q,
\end{equation}
\begin{equation}\label{current4}
J^{\dag\alpha\beta\lambda}_{2,-,3/2}=-i\frac{1}{\sqrt{2}}T^{\alpha\beta,\mu\nu}\overline{h}_{v}\gamma_{5}\gamma_{t\mu}D_{t\nu}q,
\end{equation}
and
\begin{equation}\label{current5}
J^{\dag\alpha\beta}_{2,-,5/2}=-\sqrt{\frac{5}{6}}T^{\alpha\beta,\mu\nu}\overline{h}_{v}\gamma_{5}
(D_{t\mu}D_{t\nu}-\frac{2}{5}D_{t\mu}\gamma_{t\nu}\rlap/D_{t})q ,
\end{equation}
\begin{equation}\label{current6}
J^{\dag\alpha\beta\lambda}_{3,-,5/2}=-\frac{1}{\sqrt{2}}T^{\alpha\beta\lambda,\mu\nu\sigma}\overline{h}_{v}\gamma_{t\mu}
D_{t\nu}D_{t\sigma}q ,
\end{equation}
which correspond to Eq. (\ref{current1 }), and corresponding to Eq.
(\ref{current2 }) are
\begin{equation}\label{current7}
J^{\dag\alpha}_{1,-,3/2}=-\sqrt{\frac{3}{4}}\overline{h}_{v}(D^{\alpha}_{t}-\frac{1}{3}\gamma^{\alpha}_{t}\rlap/D_{t})\rlap/D_{t}q,
\end{equation}
\begin{equation}\label{current8}
J^{\dag\alpha\beta\lambda}_{2,-,3/2}=-\frac{1}{\sqrt{2}}T^{\alpha\beta,\mu\nu}\overline{h}_{v}\gamma_{5}\gamma_{t\mu}D_{t\nu}\rlap/D_{t}q,
\end{equation}
and
\begin{flushleft}
\begin{equation}\label{current9}
J^{\dag\alpha\beta}_{2,-,5/2}=-\sqrt{\frac{5}{6}}T^{\alpha\beta,\mu\nu}\overline{h}_{v}\gamma_{5}
(D_{t\mu}D_{t\nu}-\frac{2}{5}D_{t\mu}\gamma_{t\nu}\rlap/D_{t})(-i)\rlap/D_{t}q,
\end{equation}
\begin{equation}\label{current10}
J^{\dag\alpha\beta\lambda}_{3,-,5/2}=-\frac{1}{\sqrt{2}}T^{\alpha\beta\lambda,\mu\nu\sigma}\overline{h}_{v}\gamma_{t\mu}
D_{t\nu}D_{t\sigma}(-i)\rlap/D_{t}q ,
\end{equation}
\end{flushleft}
where $h_{v}$ is the generic velocity-dependent heavy quark
effective field in HQET and $q$ denotes the light quark field. The
tensors $T^{\alpha\beta,\mu\nu}$ and
$T^{\alpha\beta\lambda,\mu\nu\sigma}$ are used to symmetrize indices
and are given by \cite{Dai}
\begin{equation}\label{tensor1}
T^{\alpha\beta,\mu\nu}=\frac{1}{2}(g^{\alpha\mu}_{t}g^{\beta\nu}_{t}
+g^{\alpha\nu}_{t}g^{\beta\mu}_{t})-\frac{1}{3}g^{\alpha\beta}_{t}g^{\mu\nu}_{t},
\end{equation}
\begin{eqnarray}\label{tensor2}
T^{\alpha\beta\lambda,\mu\nu\sigma}&=&\frac{1}{6}(g^{\alpha\mu}_{t}g^{\beta\nu}_{t}
 g^{\lambda\sigma}_{t}+g^{\alpha\mu}_{t}g^{\beta\sigma}_{t}g^{\lambda\nu}_{t}+
 g^{\alpha\nu}_{t}g^{\beta\mu}_{t}g^{\lambda\sigma}_{t}+g^{\alpha\nu}_{t}g^{\beta\sigma}_{t}g^{\lambda\mu}_{t}
 +g^{\alpha\sigma}_{t}g^{\beta\nu}_{t}g^{\lambda\mu}_{t}+
 g^{\alpha\sigma}_{t}g^{\beta\mu}_{t}g^{\lambda\nu}_{t}) \nonumber\\
 &-&\frac{1}{15}
 (g^{\alpha\beta}_{t}g^{\mu\nu}_{t}g^{\lambda\sigma}_{t}+g^{\alpha\beta}_{t}g^{\mu\sigma}_{t}g^{\lambda\nu}_{t}+
 g^{\alpha\beta}_{t}g^{\nu\sigma}_{t}g^{\lambda\mu}_{t}+g^{\alpha\lambda}_{t}g^{\mu\nu}_{t}g^{\beta\sigma}_{t}
 +g^{\alpha\lambda}_{t}g^{\mu\sigma}_{t}g^{\beta\nu}_{t}\nonumber\\
 &+&g^{\alpha\lambda}_{t}g^{\nu\sigma}_{t}g^{\beta\mu}_{t}
 +g^{\beta\lambda}_{t}g^{\mu\nu}_{t}g^{\alpha\sigma}_{t}+g^{\beta\lambda}_{t}g^{\mu\sigma}_{t}g^{\alpha\nu}_{t}
 +g^{\beta\lambda}_{t}g^{\nu\sigma}_{t}g^{\alpha\mu}_{t}).
\end{eqnarray}
Usually the currents with derivatives of the lowest order
(\ref{current1 }) are used in the QCD sum rule approach. However,
currents with derivatives of one order higher (\ref{current2 }) are
also used in some conditions because in the nonrelativistic quark
model there is a corresponding relation between the orbital angular
momenta and the orders of derivatives in the space wave functions.
As for the orbital D-wave mesons, which corresponding to derivatives
of order two, it is reasonable to use the currents (\ref{current5}),
(\ref{current6}), (\ref{current7}) and (\ref{current8}).

These currents have nice properties, they have nonvanishing
projection only to the corresponding states of the HQET in the
$m_{Q}\rightarrow\infty$ limit, without mixing with states of the
same quantum number but different $s_{l}$. Thus we can define
one-particle-current couplings as follows:
\begin{equation}\label{const1}
J^{P}=1^{-}:\langle
D^{*}_{1}(v,\varepsilon)|J^{\alpha}|0\rangle=f_{1}\sqrt{m_{D^{*}_{1}}}\varepsilon^{*\alpha},
\end{equation}
\begin{equation}\label{const2}
J^{P}=2^{-}:\langle D'_{2}(v,\varepsilon)|J^{\alpha\beta}|0\rangle=
f'_{2}\sqrt{m_{D'_{2}}}\varepsilon^{*\alpha\beta} ,
\end{equation}
\begin{equation}\label{const3}
J^{P}=2^{-}:\langle D_{2}(v,\varepsilon)|J^{\alpha\beta}|0\rangle=
f_{2}\sqrt{m_{D_{2}}}\varepsilon^{*\alpha\beta} ,
\end{equation}
\begin{equation}\label{const4}
J^{P}=3^{-}:\langle
D^{*}_{3}(v,\varepsilon)|J^{\alpha\beta\lambda}|0\rangle=
f_{3}\sqrt{m_{D^{*}_{3}}}\varepsilon^{*\alpha\beta\lambda} .
\end{equation}
The couplings $f_{i}$ are low-energy parameters which are determined
by the dynamics of the light degree of freedom. Since the pairs
($f_{1}$, $f'_{2}$) and ($f_{2}$, $f_{3}$) are related by the spin
symmetry, we will consider $f_{1}$ and $f_{2}$ hereafter. The decay
constants $f_{i}$ can be estimated from two-point sum rules,
therefore we list the sum rules after the Borel transformation. For
the ground-state heavy mesons, the sum rule for the correlator of
two heavy-light currents is well known. It is \cite{Hua1}
\begin{equation}\label{consrule1}
f^{2}_{-,\frac{1}{2}}e^{-2\bar{\Lambda}_{-,\frac{1}{2}}/T}=\frac{3}{16\pi^{2}}\int_{0}^{\omega_{c0}}
\omega^{2}e^{-\omega/T}d\omega-\frac{1}{2}\langle\bar{q}q\rangle(1-\frac{m^{2}_{0}}{4T^{2}}).
\end{equation}
For the $s_{l}^{P}=\frac{3}{2}^{-}$ doublet, when the currents
(\ref{current7}) and (\ref{current8}) are used, the corresponding
sum rule is :
\begin{equation}\label{consrule2}
f^{2}_{-,\frac{3}{2}}e^{-2\bar{\Lambda}_{-,\frac{3}{2}}/T}=\frac{1}{2^{8}\pi^{2}}\int_{0}^{\omega_{c1}}
\omega^{6}e^{-\omega/T}d\omega-\frac{5}{3\times2^{8}}\int_{0}^{\omega_{c1}}
\omega^{2}e^{-\omega/T}d\omega\langle\frac{\alpha_{s}}{\pi}GG\rangle.
\end{equation}
For the $s_{l}^{P}=\frac{5}{2}^{-}$ doublet, when the currents
(\ref{current5}) and (\ref{current6}) are used, the corresponding
sum rule is :
\begin{eqnarray}\label{consrule3}
f^{2}_{-,\frac{5}{2}}e^{-2\bar{\Lambda}_{-,5/2}/T}=\frac{1}{5\times2^{7}\pi^{2}}\int_{0}^{\omega_{c2}}
\omega^{6}e^{-\omega/T}d\omega-\frac{5}{3\times2^{6}}\int_{0}^{\omega_{c2}}
\omega^{2}e^{-\omega/T}d\omega\langle\frac{\alpha_{s}}{\pi}GG\rangle.
\end{eqnarray}

As we have just mentioned, for the amplitudes of the semileptonic
decays into excited states in the infinite mass limit, the only
unknown quantities in (\ref{rate1}), (\ref{rate2}), (\ref{rate3})
and (\ref{rate4}) are the universal functions $\tau_{1}(y)$ and
$\tau_{2}(y)$. In Ref. \cite{Col} the form factors $\tau_{1}(y)$ and
$\tau_{2}(y)$ were estimated through QCD sum rule by using currents
with derivatives of lower order, (\ref{current3}) to
(\ref{current6}). Considering that the corresponding relation
between the orbital angular momentum and the order of the derivative
mentioned above, we use the currents (\ref{current7}) and
(\ref{current8}) instead of (\ref{current3}) and (\ref{current4})
for the ($D_{1}^{*}$, $D'_{2}$) doublet. As for the ($D_{2}$,
$D^{*}_{3}$) doublet, we also use the currents (\ref{current5}) and
(\ref{current6}).

In order to calculate this two form factors by QCD sum rules, we
study the analytic properties of three-point correlators:
\begin{equation}\label{rule1}
i^{2}\int d^{4}xd^{4}ze^{i(k^{'}\cdot x-k\cdot
z)}\langle0|T[J^{\alpha}_{1,-}(x)
J^{\mu(v,v^{'})}_{V,A}(0)J^{\dag}_{0,-}(z)|0\rangle=\Gamma(\omega,\omega^{'},y)\mathcal
{L}^{\mu\alpha}_{V,A},
\end{equation}
\begin{equation}\label{rule2}
i^{2}\int d^{4}xd^{4}ze^{i(k^{'}\cdot x-k\cdot
z)}\langle0|T[J^{\alpha\beta}_{2,-}(x)
J^{\mu(v,v^{'})}_{V,A}(0)J^{\dag}_{0,-}(z)|0\rangle=\Gamma'(\omega,\omega^{'},y)\mathcal
{L}^{\mu\alpha\beta}_{V,A},
\end{equation}
where $J^{\mu(v,v^{'})}_{V}=h(v^{'})\gamma^{\mu}h(v)$ and
$J^{\mu(v,v^{'})}_{A}=h(v^{'})\gamma^{\mu}\gamma_{5}h(v)$. The
variables $k$($=P-m_{b}v$) and $k^{'}$($=P'-m_{c}v'$) denote
residual ``off-shell" momenta of the initial and final meson states,
respectively. For heavy quarks in bound states they are typically of
order $\Lambda_{QCD}$ and remain finite in the heavy quark limit.
$\Gamma(\omega,\omega^{'},y)$ and $\Gamma'(\omega,\omega^{'},y)$ are
analytic functions in the ``off-shell" energies $\omega=2v \cdot k$
and $\omega'=2v' \cdot k'$ with discontinuities for positive values
of these variables. They also depend on the velocity transfer $y=v
\cdot v'$, which is fixed in a physical region. $\mathcal {L}_{V,A}$
are Lorentz structures.

Following the standard QCD sum rule procedure, the calculations of
$\Gamma(\omega,\omega^{'},y)$ and $\Gamma'(\omega,\omega^{'},y)$ are
straightforward. First, we saturate Eqs.(\ref{rule1}) and
(\ref{rule2}) with physical intermediate states in HQET and find
that the hadronic representations of the correlators as follows:
\begin{equation}\label{pheno}
\Gamma_{hadron}(\omega,\omega^{'},y)=\frac{f_{-,\frac{1}{2}}f_{-,j_{l}}\tau_{i}(y)}
{(2\bar{\Lambda}_{-,\frac{1}{2}}-\omega-i\varepsilon)(2\bar{\Lambda}_{-,j_{l}}
-\omega^{'}-i\varepsilon)}+ \text{higher resonances},
\end{equation}
where $f_{-,j_{l}}$ are the decay constants defined in
Eqs.(\ref{const1}) and (\ref{const3}),
$\overline{\Lambda}_{-,j_{l}}=m_{-,j_{l}}-m_{Q}$. Second, the
functions can be approximated by a perturbative calculation
supplemented by nonperturbative power corrections proportional to
the vacuum condensates which are treated as phenomenological
parameters. The perturbative contribution can be represented by a
double dispersion integral in $\nu$ and $\nu^{'}$ plus possible
subtraction terms. So the theoretical expression for the correlator
has the form
\begin{equation}\label{theo}
\Gamma_{theo}(\omega,\omega^{'},y)\simeq\int d\nu d\nu^{'}
\frac{\rho^{pert}(\nu,\nu^{'},y)}
{(\nu-\omega-i\varepsilon)(\nu^{'}-\omega^{'}-i\varepsilon)}+
\text{subtractions}+\Gamma^{cond}(\omega,\omega^{'},y).
\end{equation}
The perturbative part of the spectral density can be calculated
straightforward. Confining us to the leading order of perturbation,
the perturbative spectral densities of the two sum rules for
$\tau_{1}(y)$ and $\tau_{2}(y)$ are
\begin{eqnarray}\label{perturb1}
&\rho_{pert}(\nu,\nu^{'},y)=
\frac{3}{2^{8}\pi^{2}}\frac{1}{(y+1)^{\frac{3}{2}}
(y-1)^{\frac{5}{2}}}\nu^{'}[(3\nu^{2}-(1+2y)(2\nu\nu'-\nu'^{2})]
\nonumber\\&\times\Theta(\nu)\Theta(\nu^{'})\Theta(2y\nu\nu^{'}-\nu^{2}-\nu^{'2}),
\end{eqnarray}
and
\begin{eqnarray}\label{perturb2}
&\rho_{pert}(\nu,\nu^{'},y)=\frac{3}{2^{8}\pi^{2}}\frac{1}{(y+1)^{\frac{7}{2}}
(y-1)^{\frac{5}{2}}}[(5\nu-12y\nu^{'}+3\nu^{'})\nu^{2}+(3\nu+\nu^{'})(2y^{2}-2y+1)\nu^{'2}]
\nonumber\\&\times\Theta(\nu)\Theta(\nu^{'})\Theta(2y\nu\nu^{'}-\nu^{2}-\nu^{'2}).
\end{eqnarray}
Following the arguments in Refs. \cite{Neu,Blo}, the perturbative
and the hadronic spectral densities cannot be locally dual to each
other, the necessary way to restore duality is to integrate the
spectral densities over the ``off-diagonal" variable
$\nu_{-}=\nu-\nu^{'}$, keeping the ``diagonal" variable
$\nu_{+}=\frac{\nu+\nu^{'}}{2}$ fixed. It is in $\nu_{+}$ that the
quark-hadron duality is assumed for the integrated spectral
densities. The integration region can be expressed in terms of the
variables $\nu_{-}$ and $\nu_{+}$ and we choose the triangular
region defined by the bounds: $0\leq \nu_{+}\leq \omega_{c}$,
$-2\sqrt{\frac{y-1}{y+1}}\nu_{+}\leq \nu_{-}\leq
2\sqrt{\frac{y-1}{y+1}}\nu_{+}$. As discussed in Refs.
\cite{Blo,Neu}, the upper limit $\omega_{c}$ for $\nu_{+}$ in the
region
$\frac{1}{2}[(y+1)-\sqrt{y^{2}-1}]\omega_{c0}\leqslant\omega_{c}\leqslant\frac{1}{2}(\omega_{c0}+\omega_{c2})$
is reasonable. A double Borel transformation in $\omega$ and
$\omega^{'}$ is performed on both sides of the sum rules, in which
for simplicity we take the Borel parameters equal
\cite{Neu,Hua1,Col}: $T_{1}=T_{2}=2T$. In the calculation, we have
considered the operators of dimension $D \leq 5$ in OPE. After
adding the nonperturbative parts, we obtain the sum rules for
$\tau_{1}$ and $\tau_{2}$ as follows:
\begin{eqnarray}\label{rule3}
\tau_{1}(y)f_{-,1/2}f_{-,3/2}e^{-(\bar{\Lambda}_{-,1/2}+\bar{\Lambda}_{-,3/2})/T}&=&
\frac{1}{2^{4}\pi^{2}}\frac{1}{(1+y)^{3}}\int^{\omega'_{c}}_{0}d\nu_{+}e^{-\frac{\nu_{+}}{T}}
\nu^{4}_{+}
\nonumber\\&-&\frac{T}{3\times2^{5}}\frac{2y+3}{(y+1)^{2}}\langle\frac{\alpha_{s}}{\pi}GG\rangle,
\end{eqnarray}
\begin{eqnarray}\label{rule4}
\tau_{2}(y)f_{-,1/2}f_{-,5/2}e^{-(\bar{\Lambda}_{-,1/2}+\bar{\Lambda}_{-,5/2})/T}&=&
\frac{3}{8\pi^{2}}\frac{1}{(1+y)^{4}}\int^{\omega_{c}}_{0}d\nu_{+}e^{-\frac{\nu_{+}}{T}}
\nu^{4}_{+}
\nonumber\\&-&\frac{T}{3\times2^{4}}\frac{1}{(y+1)^{3}}\langle\frac{\alpha_{s}}{\pi}GG\rangle.
\end{eqnarray}
We also derive the sum rule for $\tau_{2}$ by using the currents
(\ref{current9}) and (\ref{current10}), which appears to be
\begin{eqnarray}\label{rule5}
\tau_{2}(y)f_{-,1/2}f_{-,5/2}e^{-(\bar{\Lambda}_{-,1/2}+\bar{\Lambda}_{-,5/2})/T}&=&
\frac{21}{5\times2^{4}\pi^{2}}\frac{1}{(1+y)^{4}}\int^{\omega_{c}}_{0}d\nu_{+}e^{-\frac{\nu_{+}}{T}}
\nu^{5}_{+}
\nonumber\\&+&\frac{T^{2}}{3\times2^{4}}\frac{4y-25}{(y+1)^{3}}\langle\frac{\alpha_{s}}{\pi}GG\rangle.
\end{eqnarray}

\section{Numerical results and discussions}
\label{sec4}We now evaluate the sum rules numerically. For the QCD
parameters entering the theoretical expressions, we take the
standard values:
$\label{qcond}\langle\overline{q}q\rangle=-(0.24)^{3}
\mbox{GeV}^{3}$, $\label{gcond}\langle\alpha_{s}GG\rangle=0.04
\mbox{GeV}^{4}$, and $\label{mcond}m^{2}_{0}=0.8 \mbox{GeV}^{2}$. In
the numerical calculations, we take $2.83\mbox{GeV}$ \cite{God,Eic}
for the mass of the $s_{l}=5/2$ doublet and $2.78\mbox{GeV}$ for the
$s_{l}=3/2$ doublet. For mass of initial $B$ meson, we use
$m_{B}=5.279\mbox{GeV}$ \cite{Pdg}.

In order to obtain information of $\tau_{1}(y)$ and $\tau_{2}(y)$
with less systematic uncertainties in the calculation, we divide the
three-point sum rules by the square roots of relevant two-point sum
rules, as many authors did \cite{Neu,Hua1,Col}, to reduce the number
of input parameters and improve stabilities. Then we obtain
expressions for the $\tau_{1}(y)$ and $\tau_{2}(y)$ as functions of
the Borel parameter $T$ and the continuum thresholds. Imposing usual
criteria for the upper and lower bounds of the Borel parameter, we
found they have a common sum rule ``window":
$0.7\mbox{GeV}<T<1.5\mbox{GeV}$, which overlaps with those of
two-point sum rules (\ref{consrule1}), (\ref{consrule2}) and
(\ref{consrule3}) (see Fig. 1). Notice that the Borel parameter in
the sum rules for three-point correlators is twice the Borel
parameter in the sum rules for the two-point correlators. In the
evaluation we have taken $2.0\mbox{GeV}<\omega_{c0}<2.4\mbox{GeV}$
\cite{Hua1,Neu}, $2.8\mbox{GeV}<\omega_{c1}<3.2\mbox{GeV}$, and
$3.2\mbox{GeV}<\omega_{c2}<3.6\mbox{GeV}$. The regions of these
continuum thresholds are fixed by analyzing the corresponding
two-point sum rules. According to the discussion in Sec. \ref{sec3},
we can fix  $\omega'_{c}$ and $\omega_{c}$ in the regions
$2.3\mbox{GeV}<\omega'_{c}<2.6\mbox{GeV}$ and
$2.5\mbox{GeV}<\omega_{c}<2.7\mbox{GeV}$. The results are showed in
Fig. 2.
\begin{figure}\begin{center}
\begin{tabular}{ccc}
\begin{minipage}{7cm} \epsfxsize=7cm
\centerline{\epsffile{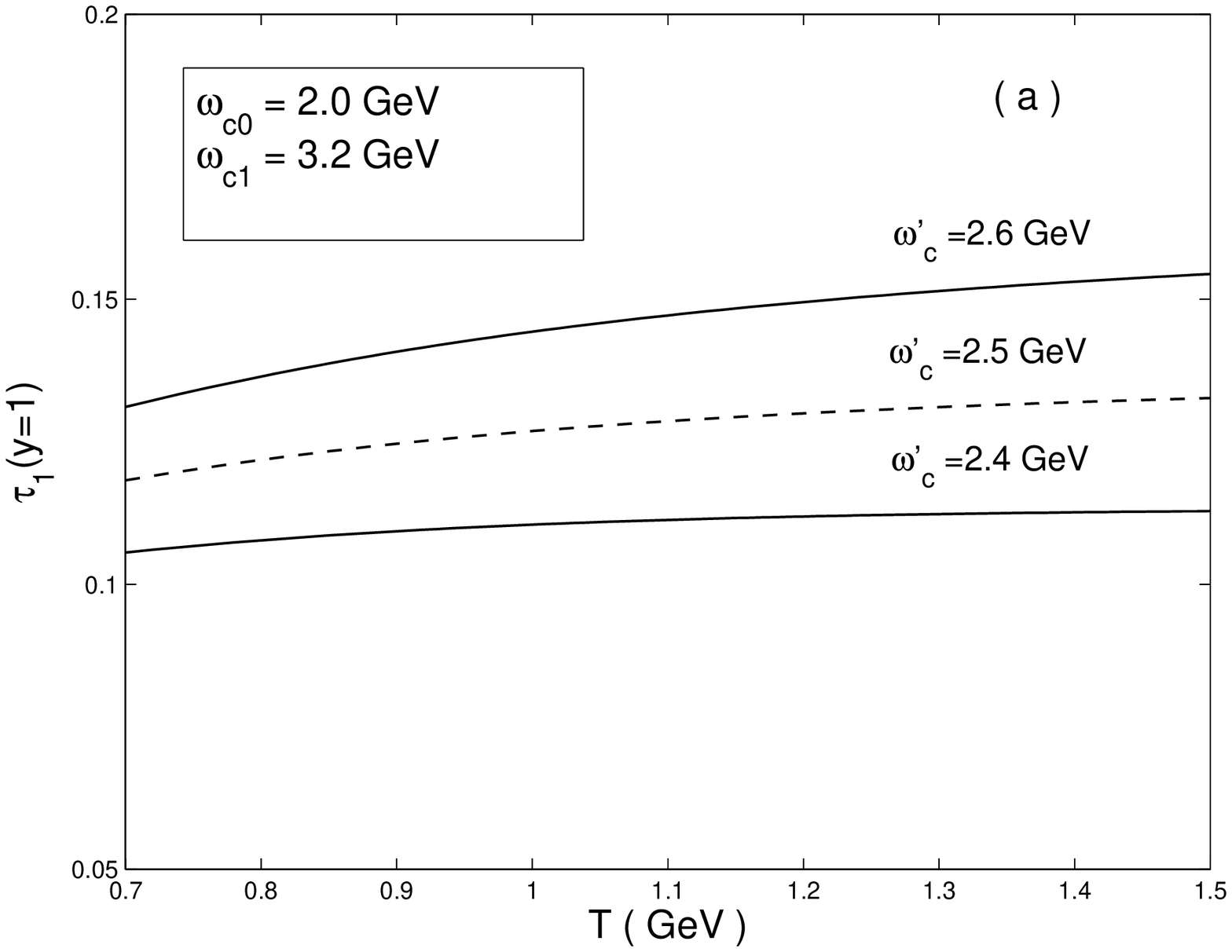}}
\end{minipage}& &
\begin{minipage}{7cm} \epsfxsize=7cm
\centerline{\epsffile{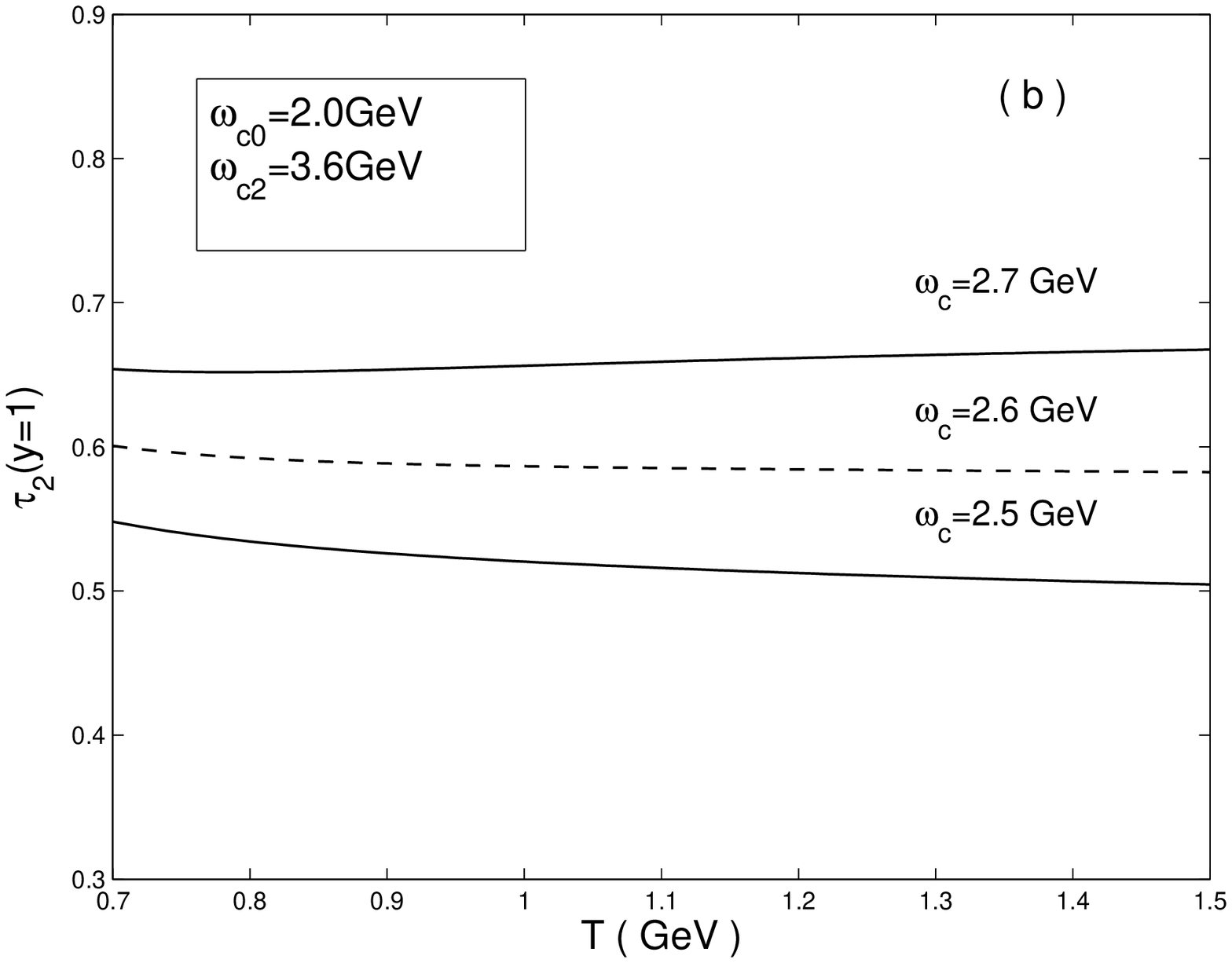}}
\end{minipage}
\end{tabular}
\caption{\it{Dependence of  $\tau_{1}(y)$ and $\tau_{2}(y)$ on Borel
parameter $T$ at $y=1$.}}
\end{center}
\end{figure}
\begin{figure}\begin{center}
\begin{minipage}{7cm}
\epsfxsize=7cm \centerline{\epsffile{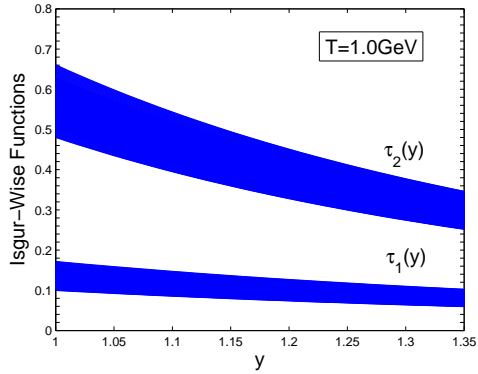}}
\end{minipage}
\caption{\it{Prediction for the Isgur-Wise functions $\tau_{1}(y)$
and $\tau_{2}(y)$.}}
\end{center}
\end{figure}
The resulting curves for $\tau_{1}(y)$ and $\tau_{2}(y)$ can be
parametrized by the linear approximation
\begin{equation}\label{linear1}
\tau_{1}(y)=\tau_{1}(1)[1-\rho^{2}_{\tau_{1}}(y-1)],\text{ }
\tau_{1}(1)=0.14\pm0.03,\text{ }\rho^{2}_{\tau_{1}}=0.13\pm0.02;
\end{equation}
\begin{equation}\label{linear2}
\tau_{2}(y)=\tau_{2}(1)[1-\rho^{2}_{\tau_{2}}(y-1)],\text{ }
\tau_{2}(1)=0.57\pm0.09,\text{ }\rho^{2}_{\tau_{2}}=0.78\pm0.13.
\end{equation}
The errors mainly come from the uncertainty due to $\omega_{c}$'s
and $T$. It is difficult to estimate these systematic errors which
are brought in by the quark-hadron duality. The maximal values of
$y$ are $y^{D^{*}_{1}}_{max} = y^{D'_{2}}_{max} =
(1+r_{1,2}^{2})/2r_{1,2}\approx 1.213$ and $y^{D_{2}}_{max} =
y^{D^{*}_{3}}_{max} = (1+r_{3,4}^{2})/2r_{3,4}\approx 1.201$. By
using the parameters $V_{cb}=0.04$,
$G_{F}=1.166\times10^{-5}\mbox{GeV}^{-2}$, we get the semileptonic
decay rates of $B\rightarrow (D_{1}^{*}, D'_{2})\ell\overline{\nu}$
and $B\rightarrow (D_{2}, D_{3}^{*})\ell\overline{\nu}$. Consider
that $\tau_{B}=1.638 \text{ps}$ \cite{Pdg}, we get the branching
ratios, respectively. All these results are listed in Table
\ref{tabe2}.
\begin{table}[h]
\caption{Predictions for the decay widths and branching ratios }
\begin{center}
\begin{tabular}{ccccccc}
\hline \hline
Decay mode & & Decay width $\Gamma$ (GeV) & & Branching ratio & & Branching ratio of Ref.\cite{Col}\\
\hline
$B\rightarrow D^{*}_{1}\ell\overline{\nu}$ & & $2.4\times10^{-18}$ & & $6.0\times10^{-6}$ & & \\

$B\rightarrow D'_{2}\ell\overline{\nu}$ & & $2.4\times10^{-18}$ & & $6.0\times10^{-6}$ & & \\

$B\rightarrow D_{2}\ell\overline{\nu}$ & & $6.2\times10^{-17}$ & & $1.5\times10^{-4}$ & & $1\times10^{-5}$ \\

$B\rightarrow D^{*}_{3}\ell\overline{\nu}$ & & $8.6\times10^{-17}$ & & $2.1\times10^{-4}$ & & $1\times10^{-5}$ \\
\hline \hline
\end{tabular}
\end{center}\label{tabe2}
\end{table}

Because of the large background from $B\rightarrow
D^{(*)}\ell\overline{\nu}$ decays, there is no experimental data
available so far. As we can see from Table \ref{tabe2}, the rates of
semileptonic $B$ decay into the $s^{P}_{l}=\frac{3}{2}^{-}$ doublet
are tiny and our results are larger than those predicted by Ref.
\cite{Col} in the $B$ to $s^{P}_{l}=\frac{5}{2}^{-}$ charmed doublet
channels. The difference comes because the way in which we choose
the parameters is different from theirs. They chose the parameters
according to other theoretical approaches. In contrast, we choose
the parameters following the way of Ref. \cite{Neu}. In addition, we
also estimate the universal form factor $\tau_{2}(y)$ with the sum
rule (\ref{rule5}) and we get almost the same result as
(\ref{linear2}). When trying to estimate the $\tau_{1}(y)$ by using
the currents (\ref{current3}) and (\ref{current4}), we find that
after the quark-hadron duality are assumed the integral over the
perturbative spectral density becomes zero. As for the $P$-wave and
the $F$-wave mesons, similar results can be obtained after the
calculations above have been carefully repeated.

The semileptonic and leptonic $B$ decay rate is about $10.9\%$ of
the total $B$ decay rate, in which the $S$-wave charmed mesons $D$
and $D^{*}$ contribute about $8.65\%$ \cite{Pdg} and the $P$-wave
charmed mesons contribute about $0.9\%$ \cite{Hua1}. Our results
then suggest that the $D$-wave charmed mesons contribute about
$0.04\%$ of the total $B$ decay rate. Sum up the branching ratios of
these semileptonic $B$ decay processes, the eight lightest charmed
mesons contribute about $9.59\%$ of the $B$ decay rate. Therefore,
semileptonic decays into higher excited states and nonresonant
multibody channels should be about $1.31\%$ of the $B$ decay rate.
Whatsoever, our result is just a leading-order estimate of the
contribution of the $D$-wave charmed mesons channels to the
semileptonic $B$ decay.

In summary, we estimate the leading-order universal form factors
describing the $B$ meson of ground-state transition into orbital
excited $D$-wave charmed resonances, the ($1^{-}$, $2^{-}$) states
($D_{1}^{*}$, $D'_{2}$), which belong to the
$s_{l}^{P}=\frac{3}{2}^{-}$ heavy quark doublet and the ($2^{-}$,
$3^{-}$) states ($D_{2}$, $D^{*}_{3}$), which belong to the
$s_{l}^{P}=\frac{5}{2}^{-}$ heavy quark doublet, by use of QCD sum
rules within the framework of HQET. The semileptonic decay widths as
well as the branching ratios we get are shown in Table \ref{tabe2}.
The predictions are larger than those predicted by Ref. \cite{Col}.
This needs future experiments for clarification. We also prove that
when $s^{P}_{l}=\frac{5}{2}^{-}$ the interpolating currents
(\ref{current1 }) and (\ref{current2 }) proposed in Ref. \cite{Dai}
are really equivalent. It is worth noting that in the estimate of
the semileptonic $B$ decay form factors when the currents
(\ref{current1 }) with quantum numbers of light degree of freedom
$s_{l}^{P}=\frac{1}{2}^{+}, \frac{3}{2}^{-}, \frac{5}{2}^{+}$ are
used for the excited charmed mesons, we find the perturbative
contributions vanish after the quark-hadron duality are assumed. In
this case we should use the currents (\ref{current2 }) which contain
derivatives of one order higher.

\begin{acknowledgments}
L. F. Gan thanks M. Zhong for useful discussions. This work was
supported in part by the National Natural Science Foundation of
China under Contract No. 10675167.
\end{acknowledgments}


\begin{thebibliography}{s2}
\bibitem{Wis}M. Neubert, Phys. Rep. 245, 259 (1994) and references therein; Aneesh V. Mannohar and Mark B. Wise, \textsl{Heavy Quark Physics}, Cambridge University Press (2000).
\bibitem{God}S. Godfrey and R. Kokoski, Phys. Rev. D 43, 1679 (1991); S. Godfrey and N. Isgur, Phys. Rev. D 32, 189 (1985).
\bibitem{Ebe}D. Ebert, V. O. Galkin and R. N. Faustov, Phys. Rev. D 57, 5663 (1998).
\bibitem{Dai}Y. B. Dai, C. S. Huang, M. Q. Huang, and C. Liu, Phys. Lett. B 390, 350 (1997); Y. B. Dai, C. S. Huang, and M. Q. Huang, Phys. Rev. D 55,
5719 (1997).
\bibitem{Neu}M. Neubert, Phys. Rev. D 45, 2451 (1992); 46, 3914 (1992).
\bibitem{Cap}S. Capstick and S. Godfrey, Phys. Rev. D 41, 2856 (1990).
\bibitem{Cve}G. Cveti\v{c} , C. S. Kim, Guo-Li Wang and Wuk Namgung, Phys. Lett. B 596, 84 (2004); Guo-Li Wang, Phys. Lett. B 633, 492 (2006).
\bibitem{Dai3}Y. B. Dai, C. S. Huang, M. Q. Huang, H. Y. Jin, and C. Liu, Phys. Rev. D 58, 094032 (1998).
\bibitem{Dai2}Y. B. Dai, C. S. Huang, and H. Y. Jin,  Z. Phys. C 60, 527-534 (1993); Phys. Lett. B 331, 174 (1994);  Y. B. Dai and H. Y. Jin, Phys. Rev. D 52, 236 (1995).
\bibitem{Eic}E. J. Eichten, C. T. Hill, and C. Quigg, Phys. Rev. Lett.71, 4116 (1993).
\bibitem{Xia}X. H. Zhong and Q. Zhao, Phys. Rev. D 78, 014029 (2007).
\bibitem{Wei}W. Wei, X. Liu, and S. L. Zhu, Phys. Rev. D 75, 014013 (2007).
\bibitem{Cle}A. Anastassov et al. (CLEO Collaboration), Phys. Rev. Lett. 80, 4127 (1998).
\bibitem{Ale}D. Buskulic et al. (ALEPH Collaboration), Phys. Lett. B 395, 373 (1997); Z. Phys. C 73, 601 (1997).
\bibitem{Aub1}B. Aubert et al. (BARBAR Collaboration), hep-ex/0808.0333.
\bibitem{Aub}B. Aubert et al. (BARBAR Collaboration), Phys. Rev. Lett. 97, 222001 (2006).
\bibitem{Bel}K. Abe et al. (BELLE Collaboration), hep-ex/0608031.
\bibitem{Wis1}A. K. Leibovich, Z. Ligeti, I. W. Stewart, and M. B. Wise, Phys. Rev. Lett.78, 3995 (1997); Phys. Rev. D 57, 308 (1998).
\bibitem{Shi}M. A. Shifman, A. I. Vainshtein, and V. I. Zakharov, Nucl. Phys. B 147, 385 (1979); 147, 448 (1979); V. A. Novikov, M. A. Shifman and A. I. Vainshtein, and V. I.
Zakharov, Fortschr. Phys. 32, 585 (1984).
\bibitem{Hua1}M. Q. Huang and Y. B. Dai, Phys. Rev. D 59, 034018 (1999); 64, 014034 (2001).
\bibitem{Ebe1}D. Ebert, R. N. Faustov and V. O. Galkin, Phys. Rev. D 61, 014016 (1999); 75, 074008 (2007).
\bibitem{Dea}A. Deandrea, N. Di Bartolomeo, R. Gatto, G. Nardulli and A. D. Polosa, Phys. Rev. D 58,
034004 (1998); V. Mor\'{e}nas, A. Le Yaouanc, L. Oliver, O. P\`{e}ne
and J. C. Raynal, Phys. Rev. D 56, 5668 (1997).
\bibitem{Fal}A. F. Falk, Nul. Phys. B 378, 79 (1992); A. F. Falk and M. Luke,  Phys. Lett. B 292, 119 (1992).
\bibitem{Col}P. Colangelo, F. De Fazio and G. Nardulli, Phys. Lett. B 478, 408 (2000).
\bibitem{Blo}B. Blok and M. Shifman, Phys. Rev. D 47, 2949 (1993).
\bibitem{Pdg}Particle Data Group, C. Amsler et al., Phys. Lett. B 667, 1 (2008).


\end{thebibliography}
\end{document}